\documentclass[conference]{IEEEtran}

\usepackage{cite}

\usepackage{graphicx}

\usepackage{algorithmic}

\usepackage[tight,footnotesize]{subfigure}

\usepackage[caption=false,font=footnotesize]{subfig}

\usepackage{stfloats}

\usepackage{url}

\usepackage{color}

\usepackage{listings}
\usepackage{subfigure}

\newcommand{\tfirst}{\raisebox{-3pt}{\includegraphics[height=0.17in]{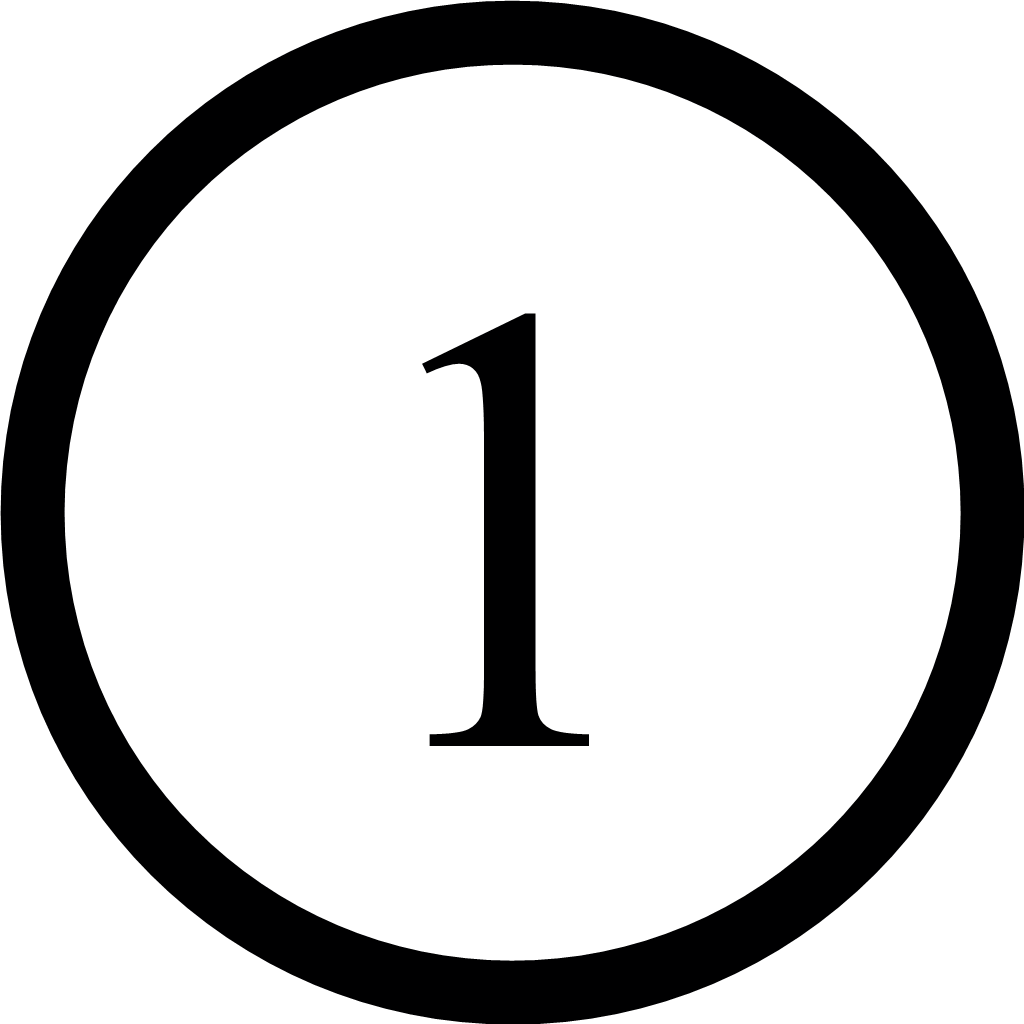}}}
\newcommand{\tseconda}{\raisebox{-3pt}{\includegraphics[height=0.17in]{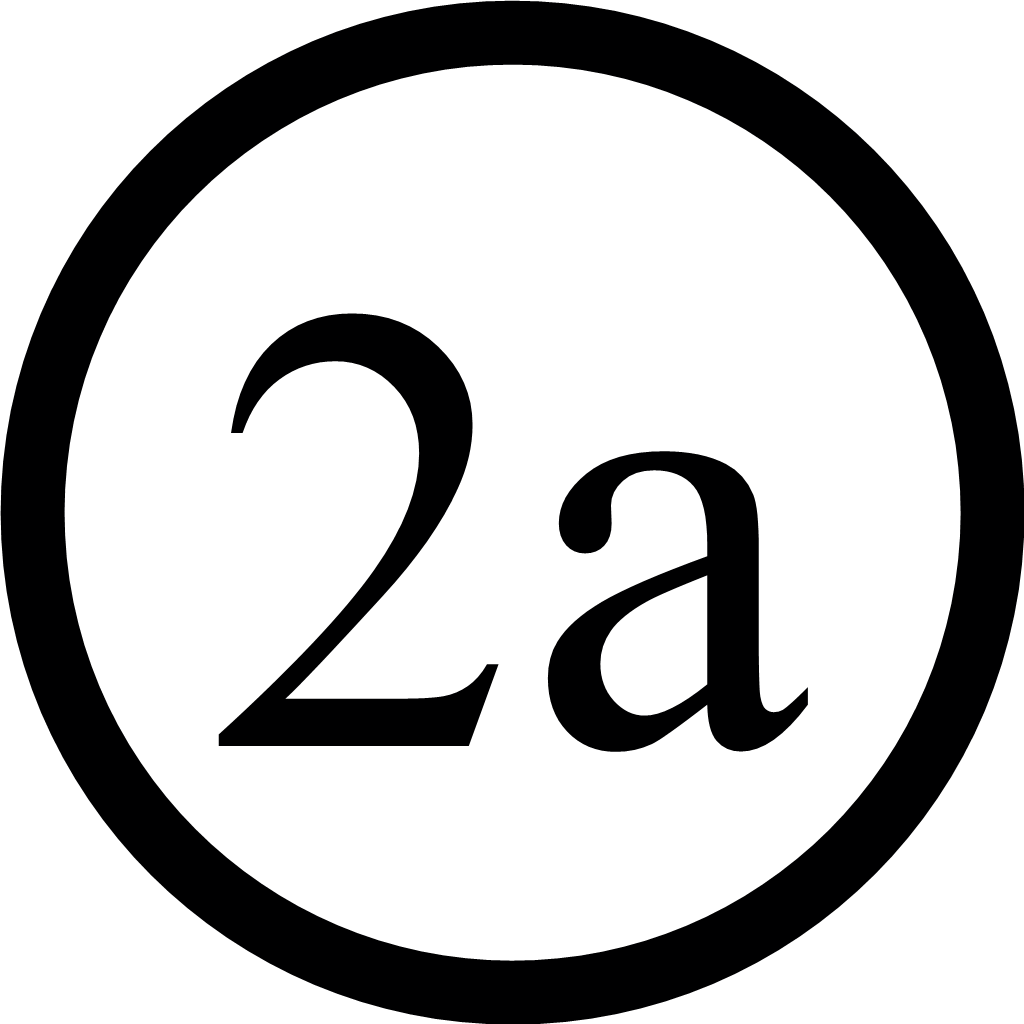}}}
\newcommand{\tsecondb}{\raisebox{-3pt}{\includegraphics[height=0.17in]{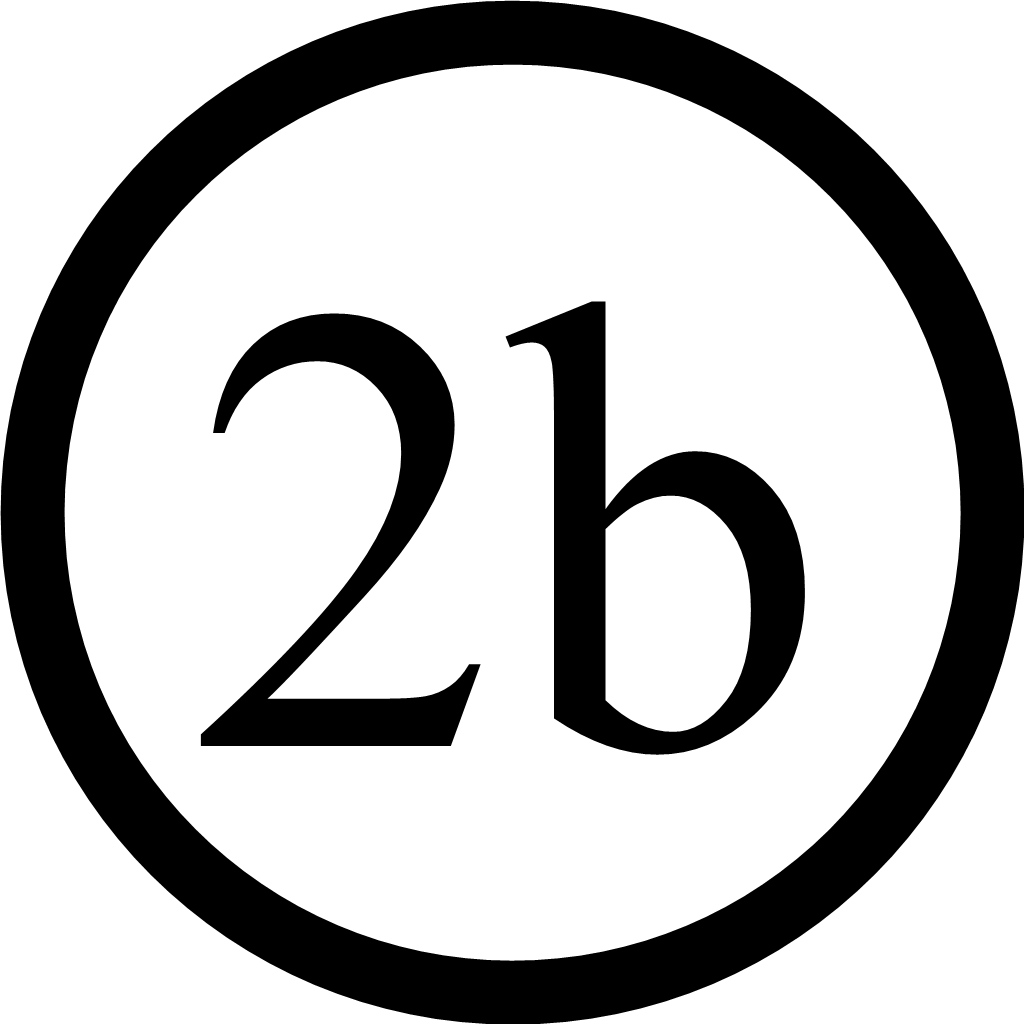}}}
\newcommand{\tthird}{\raisebox{-3pt}{\includegraphics[height=0.17in]{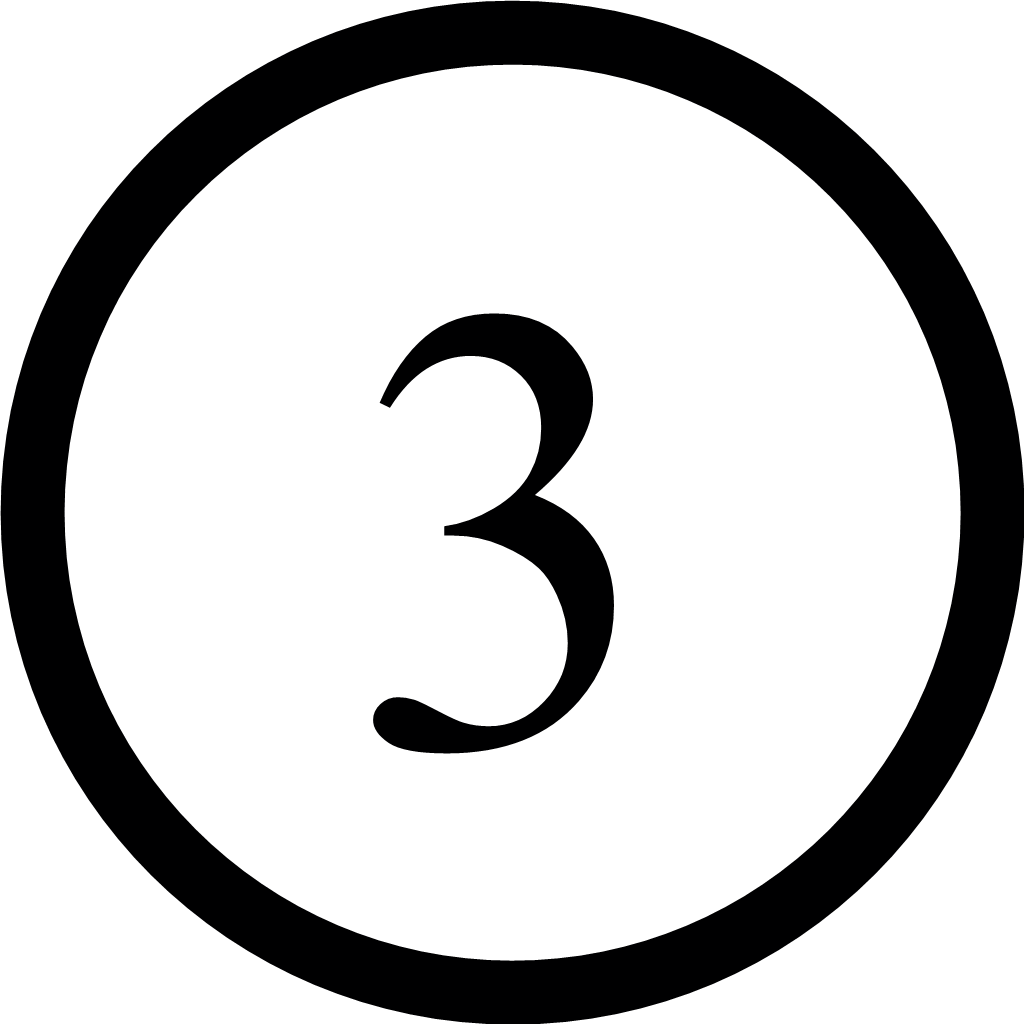}}}

\newcommand{\fitem}[1]{\item[]\hspace{-0.17in}\textbf{#1}}

\begin{document}

  \title{A Digital Marketplace Combining WS-Agreement, Service Negotiation Protocols and Heterogeneous Services}


\author{\IEEEauthorblockN{Ralph Vigne}
\IEEEauthorblockA{\textit{Faculty of Computer Science} \\
\textit{University of Vienna}\\
Vienna, Austria\\
ralph.vigne@univie.ac.at
}
\and
\IEEEauthorblockN{Juergen Mangler}
\IEEEauthorblockA{\textit{Faculty of Computer Science} \\
\textit{University of Vienna}\\
Vienna, Austria\\
juergen.mangler@univie.ac.at
}
\and
\IEEEauthorblockN{Erich Schikuta}
\IEEEauthorblockA{\textit{Faculty of Computer Science} \\
\textit{University of Vienna}\\
Vienna, Austria \\
erich.schikuta@univie.ac.at}
}

\maketitle

\begin{abstract}
With the ever increasing importance of web services and the Cloud as a reliable
commodity to provide business value as well as consolidate IT infrastructure,
electronic contracts have become very important. WS-Agreement has itself
established as a well accepted container format for describing such contracts.
However, the semantic interpretation of the terms contained in these contracts,
as well as the process of agreeing to contracts when multiple options have to
be considered (negotiation), are still pretty much dealt with on a case by case
basis. In this paper we address the issues of diverging contracts and varying
contract negotiation protocols by introducing the concept of a contract aware
marketplace, which abstracts from the heterogeneous offers of different
services providers. This allows for the automated consumption of services
solely based on preferences, instead of additional restrictions such as
understanding of contract terms and/or negotiation protocols. We also
contribute an evaluation of several existing negotiation concepts/protocols. We
think that reducing the complexity for automated contract negotiation and thus
service consumption is a key for the success of future service and Cloud
infrastructures.
\end{abstract}
\IEEEpeerreviewmaketitle

\section{Introduction}
\label{sec:introduction}

Over the last years, WS-Agreement (WSAG)
\cite{ludwig_ws-agreement_2006,andrieux_web_2004} has become a broadly used
standard in the research field of Grid and Cloud computing for providing
electronic contracts \cite{broberg_market-oriented_2008,
sakellariou_flexibility_2005, haq_framework_2009, zulkernine_adaptive_2011}.
Their success is based on a flexible means to (1) define involved parties, (2)
define the context of agreements, and (3) provide necessary terms for
guarantees. This makes them observable and consumable and therefore enables
automated service integration in Grid and Cloud based systems (see e.g.
\cite{yan_autonomous_2007, ludwig_cremona:_2004}).

While the flexibility of WSAG allowed it to become the de-facto standard for
defining electronic contracts, there is no similar standard to describe
negotiation protocols. Research defines automated service negotiation (e.g.
\cite{venugopal_negotiation_2008, zulkernine_adaptive_2011,
comuzzi_architecture_2005, parkin_framework_2006,
hasselmeyer_implementing_2007, pichot_dynamic_2008, faratin_negotiation_1998})
as the process of establishing business relations between service providers and
consumers. It builds on the fact that services providers can guarantee
different levels of service to consumers (traditionally: bandwidth, cpu,
\ldots). These service properties of course (1) have effects on each other, (2)
lead to different prices, and (3) have to be matched to the preferences of
service consumers. FIPA already standardized some custom negotiation protocols
(e.g. \cite{odell_fipa_2002}) for the Grid / Cloud communities. These protocols
often are very specific to their respective application domains, i.e.  dealing
with a specific set of properties, how to balance them, and how to match them
to user preferences in order to maximize the benefit for all participants.
However, for all these approaches, providers and consumers have to implement
and conform to the negotiation protocol. A generic way to describe complex
protocols is still missing.

The contribution of this paper is a means to integrate arbitrary negotiation
protocols and enable service domains to leverage WSAG-based electronic
contracts. The vision is to allow service consumers to participate in
negotiations solely based on their preferences, not on their understanding of
the negotiation protocol, thus allowing for the coexistence of negotiation
protocols, instead of their standardization.

In order to achieve this feat we built on the Marketplace presented in Vigne et
al. \cite{vigne_structured_2012}. The basic idea behind this Marketplace is to
provide a repository of heterogeneous services (different APIs, protocols) and
a set of microflows (small code snippets in a process language like e.g. BPEL).
While similar approaches (semantic repositories; UDDI+OWLS) concentrate on
finding services through semantic annotations, the Marketplace focuses on
standardizing the interaction with services for particular application domains.
We further improve on this marketplace concept by \ldots

\begin{enumerate}

  \item Utilizing the Marketplace information and conventions in order to
  automatically generate contracts. This allows participating service providers
  to benefit from WSAG without the burden of building WSAG handling and
  negotiation into their services.

  \item Explaining how to facilitate the basic concepts of the Marketplace in
  order to standardize electronic contracts for similar services from different
  service providers.

  \item Explaining how to use the concept of microflows (see above) in order to
  realize bilateral (one-to-one) and multilateral (one-to-many) automated
  service negotiation. The goal is to abstract from particular services, and
  instead allow service consumers to participate in a negotiation solely based
  on their preferences, not based on their knowledge of negotiation protocols.

\end{enumerate}

We further provide an evaluation of the viability of our approach by means of a
case study. This case study consists of a categorization of analysis of current
Grid and Cloud negotiation protocols and how they can be realized in the
Marketplace framework.

Our research in this area is based on our experience with large scale data stores~\cite{schikuta1998vipios} and complex applications in Grids and Clouds~\cite{mach2012generic,schikuta2004n2grid,cs745}, and strongly motivated by our focus on Web-based workflow optimizations~\cite{schikuta2008grid,kofler2009parallel} and their respective management~\cite{stuermer2009building}.

In Section \ref{sec:example} we provide a short recap about the used
Marketplace and its provided concepts and functionality. In Section
\ref{sec:related} we discuss related work from the fields of WSAG negotiation
for Grid- and Cloud-based systems.  How WSAG handling is integrated into the
Marketplace is discussed in Section \ref{sec:agreements}. A case study about
bilateral and multilateral negotiation protocols and additionally an example
for strategic negotiation is given in Section \ref{sec:negotiation}.  In
Section \ref{sec:conclusion} we conclude this work and provide an outlook for
future work.

\section{The Marketplace}
\label{sec:example}

As this work is based on the Marketplace presented by Vigne et al.
\cite{vigne_structured_2012}, it is essential for further discussions to have a
basic understanding of the implemented concepts and functionality.

\begin{figure}[ht]
  \begin{center}
     \includegraphics[width=0.9\columnwidth]{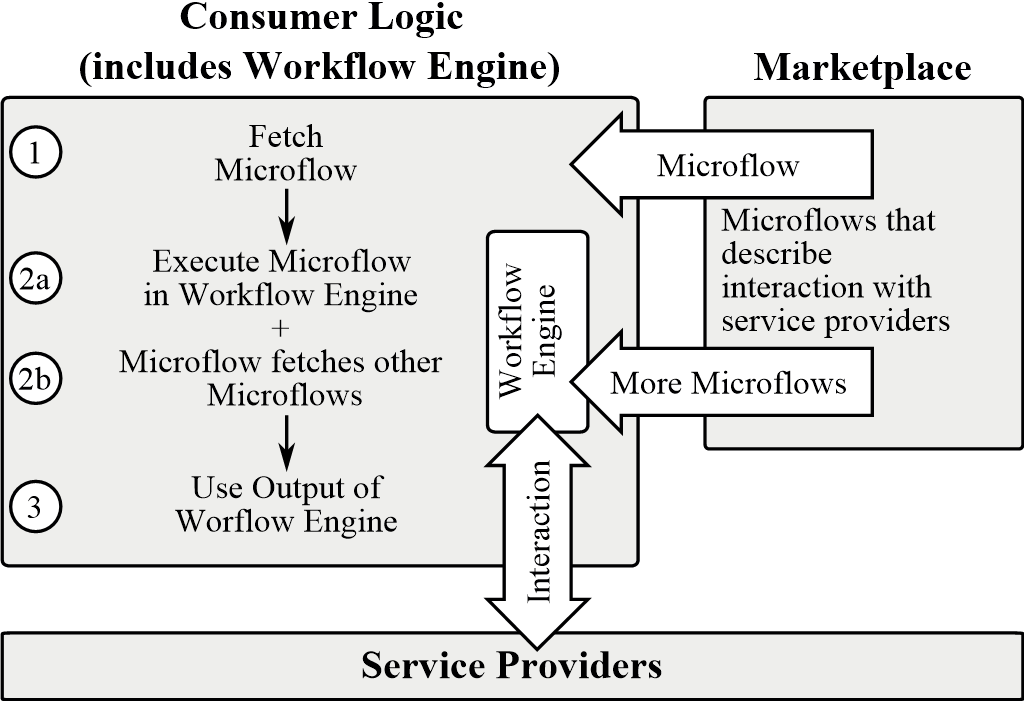}
  \end{center}
  \vspace{-10pt}
  \caption{The Marketplace}
  \label{fig:marketplace}
\end{figure}

As can be seen in Figure \ref{fig:marketplace}, the Marketplace holds a set of
microflows that describe \textbf{how to interact with services}. The
Marketplace is strictly \textbf{passive}, connotes it does not act as a
middleware or proxy. Its purpose is to allow a service consumer to implement a
consumer logic against a standardized API for a particular application
\textit{domain}.

There are two kinds of microflows contained in the Marketplace:

\begin{itemize}

  \item Service microflows that implement an interaction with a particular
  service regarding the requirements of the domain API. This group of
  microflows is from now on referred to as \textbf{instance level operations}.

  \item Domain microflows that implement logic to fetch and invoke service
  microflows to select particular services for particular inputs. Each
  microflow correlates with one operation in the domain API. Thus we will
  further on refer to this group of microflows as \textbf{class level
  operations}.

\end{itemize}

From the service consumer's point of view, the process of doing business with
service providers that participate in the Marketplace is totally transparent:

\begin{itemize}

  \fitem{\tfirst} The consumer logic fetches the microflow that represent a
  particular \textbf{class level operation} from the Marketplace.

  \fitem{\tseconda} The consumer logic instantiates a Workflow Engine (WFE)
  with the microflow, and invokes the microflow with a set of input parameters
  (the required input parameters are also defined at the class level, e.g. as a
  contract).

  \fitem{\tsecondb} While the class level microflow is running, it fetches and
  executes further \textbf{instance level operations} from the Marketplace, in
  order to find the service provider best matching the consumer's preferences
  and use it.  This (potentially complex) selection of services is furthermore
  called \textbf{negotiation}, and is one of the subjects of this paper. An
  instance level microflow (also executed by the WFE) directly interacts with a
  service, and transforms / filters information to make it usable for class
  level microflows.

  \fitem{\tthird} When a particular class level microflow is finished, its
  results can be used as part of the consumer logic.

\end{itemize}

For this paper we again reuse the Cinemas example presented in
\cite{vigne_structured_2012}. The cinema domain consist of three {class level
operations}: (1) \textbf{Search} to search for shows of a particular movie, (2)
\textbf{Book} to book tickets for a particular show and (3) \textbf{Search \&
Book} which combines the two (to demonstrate a complex, multi-part
negotiation).

\section{Related Work}
\label{sec:related}

As mentioned before, WS-Agreements have become the de-facto standard for
electronic contracts in the web service context. In this section we give an
overview of WSAG and autonomous Service Level Agreement (SLA) negotiation
related research.

Since Andrieux et al. specified WSAG \cite{andrieux_web_2004} in 2004 it has
become a widely used standard in the area of web services to define contracts
between two or more parties. Inside a WSAG, all agreed information is
structured into different parts. The main parts are namely:

\begin{enumerate}

  \item \textbf{Context:} This part contains the meta-data of the agreement
  e.g. participant's and life-time.

  \item \textbf{Terms:} This part contains data about the agreement itself. It
  consists of \textbf{Service Terms}, which represent the functionality that
  will be delivered (so called Items), and the \textbf{Guarantee Terms}, which
  define assurances on the service quality associated with the defined Service
  Terms.

\end{enumerate}

This formalized and comprehensive way of describing agreements and the
according guarantees make WSAG \textbf{automatically computable and observable}
\cite{ludwig_cremona:_2004, oldham_semantic_2006}. Sakellariou et al.
\cite{sakellariou_flexibility_2005} extended Guarantee Terms to be represented
as the \textbf{result of a function} instead of static values and therefore
making it more flexible when observing complex ones.

As WSAG is originally designed for SOAP-based services, Kuebert et al.
\cite{kuebert_restful_2011} introduced a way to use them also for
\textbf{RESTful services} which increased in numbers rapidly over the last few
years.

Haq identifies in \cite{haq_framework_2009} the \textbf{compensation of the
service consumer's high dependency on the service provider}, as a result of
using WSAG, as an additional reason for their common use in the field of Grid
and Cloud computing. He further states that the emerging of Composite Service
Providers leads to \textbf{complex value chains} and therefore consist of
aggregated WSAG. While our approach provides a simple concatenation of these
WSAGs, he states that not every information, created during the composition,
should be exposed to each participating party. Therefore he introduced
\textbf{SLA-Views}, which are basically a customized view of the aggregated
WSAG for each participating party.

Broberg et al.  \cite{broberg_market-oriented_2008} gives an overview about
\textbf{state-of-the-art and future directions} in the field of market-based
Grid and Utility computing. Although they focus on negotiation strategies
related to scheduling problems, they additionally conclude that it is important
to overcome the restrictions in flexibility of negotiation protocols to exploit
the benefits for service consumers generated by a market with competing service
providers.

Zulkernine et al. \cite{zulkernine_adaptive_2011, zulkernine_policy-based_2009}
developed a flexible system for WSAG negotiation with the focus on
\textbf{multilateral negotiation strategies} based on decision support systems.
They also conclude that a way to define flexible and provider specific
protocols would further increase the quality of negotiation results.

Faratin et al. \cite{faratin_negotiation_1998} give a detailed discussion how
such sets of decisions and preferences could look like. Within their example
implementation of \textbf{autonomous service negotiation}, different offers are
evaluated and counteroffers are created on the base of a value scoring
algorithm. They define \textbf{negotiation tactic} as the short-term decision
making focused on a narrow set of criteria while the \textbf{negotiation
strategy} represents the weighting of criteria over time.

As more complex negotiation protocols come into focus of research, WSAG reaches
its limitation as it offers \textbf{only two messages}, namely \textit{offer}
(as an input for negotiation) and \textit{agreement} (as the output of an
negotiation) and therefore makes it only feasible for the ``Contract Net
Interaction Protocol'' \cite{smith_contract_1980} (see also Figure
\ref{fig:negotiation:contract_net}).  Hung et al. elaborated that a formalized
negotiation must consist of three different parts which they further used as
groundwork when developing the ``WS-Negotiation Specification''
\cite{hung2004ws}.  These three parts are namely:

\begin{enumerate}

  \item \textbf{Negotiation Message} describes the messages which are sent and
  received during a negotiation. These messages can be of different types,
  namely \textit{Offer, Counteroffer, Rejected, Accepted, Expired,
  SinglePartySigned, Signed,} and \textit{Unsigned}.

  \item \textbf{Negotiation Protocol} defines a set of rules and mechanisms the
  negotiating parties must follow. It uses {negotiation primitives} to
  describe different interactions between the parties and what pre- and
  post-conditions are associated to each primitive (similar to declarative
  workflow descriptions). The following primitives are defined: \textit{Call
  for Proposal, Propose, Accept, Terminate, Reject, Acknowledge, Modify,} and
  \textit{Withdraw}.

  \item \textbf{Negotiation Decision Making} represents an internal and private
  decision making process based on a cost-benefit function and therefore
  represents the \textit{negotiation strategy}.

\end{enumerate}

Venugopal et al. \cite{venugopal_negotiation_2008} focused on the negotiation
protocol and proposed Rubinstein's approach of ``Alternating Offers''
\cite{rubinstein_perfect_1982} (see also Figure
\ref{fig:negotiation:alternate_offers}) as one way to increase the quality of
\textbf{bilateral} service negotiation. A \textbf{multilateral} adaptation of
this  protocol has been standardized by \cite{odell_fipa_2002}. Yan et al.
\cite{yan_autonomous_2007} further extended this protocol (see also Figure
\ref{fig:negotiation:competing_resources}) because of its shortcomings in
\textbf{negotiation coordination} i.e. coordinating multiple counteroffers for
various service providers.

\section{WS-Agreements \& The Marketplace}
\label{sec:agreements}

As stated in the introduction, we try to reuse the Marketplace to commoditize
the utilization of contracts (in our case WS-Agreements (WSAG)). We envision
this to work similar to the way the Marketplace deals with heterogeneous
services (i.e.  same application domain but different APIs, slightly different
semantics).

\subsection{Concepts} 
\label{sec:agreements:concepts}

For services in the same application domain, it is possible to provide a unified
contract template \cite{altmann_vieslaf_2009}, that covers possible terms and
restrictions for all services, but has a certain characteristic for particular
services. We again propose a split analog to the class level / instance level
differentiation described in Section \ref{sec:example}:

\begin{itemize}

  \item By providing domain contracts that define the interaction with the
  Marketplace, as well as common criteria for selecting any service from the
  domain.

  \item By providing service contracts that all have the same structure and
  define the level of service (quality) that can be expected from particular
  services.

\end{itemize}

Contracts are not some information that is fetched from a repository, but they
are something that is tightly coupled to the usage of services. They are
created, negotiated, and accepted while using certain class level operations of
the Marketplace. Contracts may of course differ from customer to customer, and
are strongly depending on the kind of service that a service consumer tries to
utilize (i.e. the input parameters).

Thus  before the consumer logic can execute certain operations from the
Marketplace (i.e. fetch microflows and execute them in a WFE as describe in
Section \ref{sec:example}), it has to ensure that the properties of a service
match its preferences. This can be done by means of a WSAG.

The typical effort for utilizing WSAG includes (1) fetching a contract
template, (2) making an offer, and (3) receiving the agreement (or rejection).
Of course additional logic may be necessary to (a) ponder which services to
actually make an offer, and (b) decide between multiple possible agreements
(under the condition that it is not necessary to enact on all agreements).

Furthermore we have to consider that services may know nothing about WSAG, by
still offer an API to achieve something similar. We conclude: it will be
necessary to (1) transform WSAG, or (2) implement logic to create WSAG
templates and / or decide if a service accepts an offer. This logic can, as
explained in Section \ref{sec:example} be expressed in microflows at service
level.

\begin{figure}[ht]
  \begin{center}
     \includegraphics[width=0.9\columnwidth]{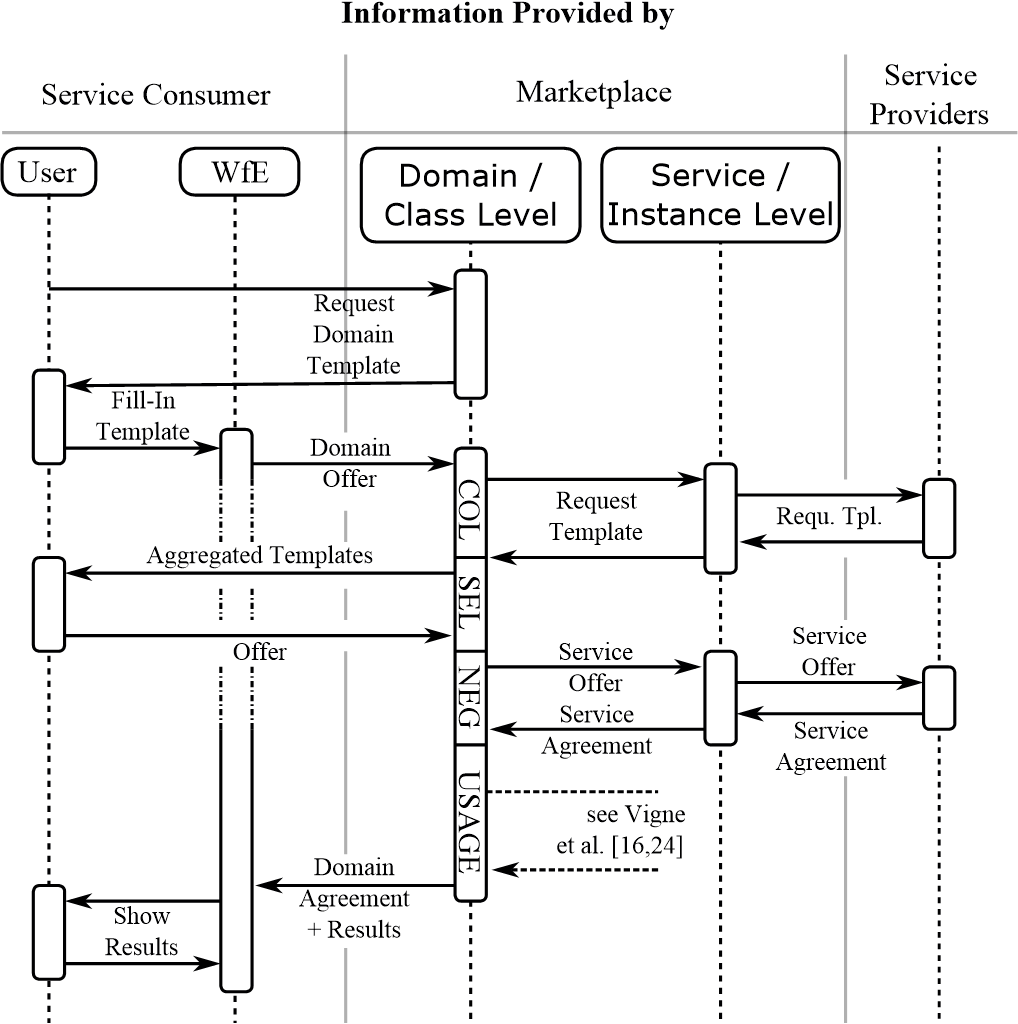}
  \end{center}
  \vspace{-10pt}
  \caption{Abstract Communication Overview of Using the Marketplace}
  \label{fig:communication}
\end{figure}

For Figure \ref{fig:communication} keep in mind, that the Marketplace is no
middleware. It just depicts which kind of information have to be fetched by the
WFE, and which kind of information has to be provided by the user and services
in order to enable WSAG and SLA negotiation.

As illustrated, the service consumer, after having received a domain level
template, a user has to provide the necessary information to create a
\textit{domain offer} i.e. WSAG offer. This domain offer is used as input for
the according domain operation (microflow enacted by the WFE).

First the operation \textit{collects (COL)} the \textit{templates} of all
service providers.  This is done executing the according \textit{class and
succeeding instance level microflows}. When all templates are collected, they
are filtered using the information provided via the \textit{domain offer}.

Next, these \textit{templates are aggregated} and presented to the user who
\textit{selects (SEL)} according to personal preferences. Now the selected
templates are supplemented with information included in the \textit{domain
offer} and therefore become \textit{service offers}.

During the \textit{negotiation (NEG)} this \textit{service offers} are used to
negotiate with the according service providers (see Section
\ref{sec:negotiation} for details about negotiation protocols) using a
negotiating class level operation (microflow) and in succession the specific
instance level operation. At the end of the negotiation, a \textit{service
agreement}, authorizing for the usage of the service, is defined.

Providing this \textit{service agreement}, the service consumer's WFE can
\textit{use the service (USAGE)}. How this is done is discussed in detail in
\cite{vigne_structured_2012}.

After the service is used, a \textit{domain agreement} is created, using among
others the \textit{service agreement} received during the negotiation.  This
\textit{domain agreement} and the \textit{execution results} of the service
execution, representing the defined output of the domain operation, are made
persistent for the user.

\subsection{Implementation} 

The implementation of a WSAG aware system requires utilizing existing
information from the Marketplace about the (1) included services (their
properties) and (2) their usage.

\begin{itemize}

  \item Service properties are a set of static attributes that each service
  provider must provide to participate in a certain domain of the Marketplace.
  Examples for the cinema domain are address, number of cinemas/seats, smoking
  allowed, food corner available, \ldots Service properties lead to simple WSAG
  guarantee terms, with a particular unit and/or value range, that may be
  observable from the outside. Also these properties can be used to create the
  WSAG business values.

  \item Required input to class level operations (which is transformed to
  suitable input for services in instance level operations). Examples include
  the actual seat to book. The presence of such input parameters may be
  required or optional, and again expressed in particular units or as a value
  range.  Thus such input parameters lead to WSAG qualifying conditions and/or
  creation constraints (how to fill out a WSAG template in order to create a
  valid offer).

  \item Output of class level operations are the most important part. For them
  the main purpose of WSAG, to monitor the quality of results, fully applies.
  They lead to WSAG guarantee terms with a particular unit and/or value range.
  The outside observability may of course be very different across a range of
  services. Thus it has to be linked to a class level operation, that in turn
  may invoke monitoring functionality at the instance level (e.g.  Sakellariou
  et al. \cite{sakellariou_flexibility_2005}).

\end{itemize}

The creation of a WSAG \textit{domain template} is strictly internal, and fully
transparent for the Marketplace consumers. For its implementation we are
building on the existing mechanisms of the Marketplace, mainly the definition
of properties, and input / output schemas.

\begin{lstlisting}[caption=Properties Schema,label=lst:properties]
<rng:grammar xmlns:datatypeLibrary="..." xmlns:rng="..." xmlns:prop="..." xmlns:wsag="..."> 
  <rng:start> 
    <prop:properties xmlns:prop="..." xmlns:d="..."> 
      <rng:element name="prop:address"> 
        <d:caption xml:lang="en">Address</d:caption> 
        <d:caption xml:lang="de">Adresse</d:caption> 
        <rng:element name="prop:ZIP" wsag:item="Zip">
          <d:caption xml:lang="en">ZIP Code</d:caption> 
          <d:caption xml:lang="de">PLZ</d:caption> 
          <rng:data type="integer"><!-- 4 digits -->
            <param name="minInclusive">1000</param>
            <param name="maxInclusive">9999</param>
          </rng:data>
          <wsag:extension>
            <!-- specific WSAG information -->
          </wsag:extension>
        </rng:element> 
        ...
      </rng:element> 
    </prop:properties> 
    <wsag:general>
      <!-- 
        General information about the WS-A template 
      -->
    </wsag:general>
  </rng:start> 
</rng:grammar>
\end{lstlisting}

As can be seen in Listing \ref{lst:properties}, we added a WSAG specific
section to the already existing XML RNG Schema, to allow for WSAG specific
information to be added (\textbf{wsag:item}). As properties are class level
information, it is up to a service domain expert to shape this basis for all
contracts.

\begin{lstlisting}[caption=Class Level Input Message Schema,label=lst:input]
<rng:grammar xmlns:datatypeLibrary="..." xmlns:rng="..."> 
  <rng:start> 
    ...
    <rng:element name="input-message"> 
      ...
      <rng:element xmlns:d="..." name="date" wsag:Item="DateOfShow"> 
        <d:caption xml:lang="en">Date</d:caption> 
        <d:caption xml:lang="de">Datum</d:caption> 
        <rng:data type="date"/> 
        <wsag:extension>
          <!-- specific WSAG information -->
        </wsag:extension>
      </rng:element> 
      <rng:element xmlns:d="..." name="seats" wsag:Item="NumberOfSeats"> 
        <d:caption xml:lang="en">Seats</d:caption> 
        <d:caption xml:lang="de">Sitze</d:caption> 
        <rng:data type="integer"> 
          <param name="minInclusive">1</param>
          <param name="maxInclusive">6</param>
        </rng:data>
        <wsag:extension>
          <!-- specific WSAG information -->
        </wsag:extension>
      </rng:element> 
    </rng:element> 
    <wsag:general>
      <!-- 
        General information about the WSAG template 
      -->
    </wsag:general>
  </rng:start> 
</rng:grammar>
\end{lstlisting}

In Listing \ref{lst:input} an example class level input schema can be seen.
Again it is possible to reuse datatype related information, but additionally
units and/or expected ranges could be further described in WSAG. 

\subsection{Aggregation}
\label{sec:agreements:aggregation}

\begin{figure}[ht]
  \begin{center}
     \includegraphics[width=0.7\columnwidth]{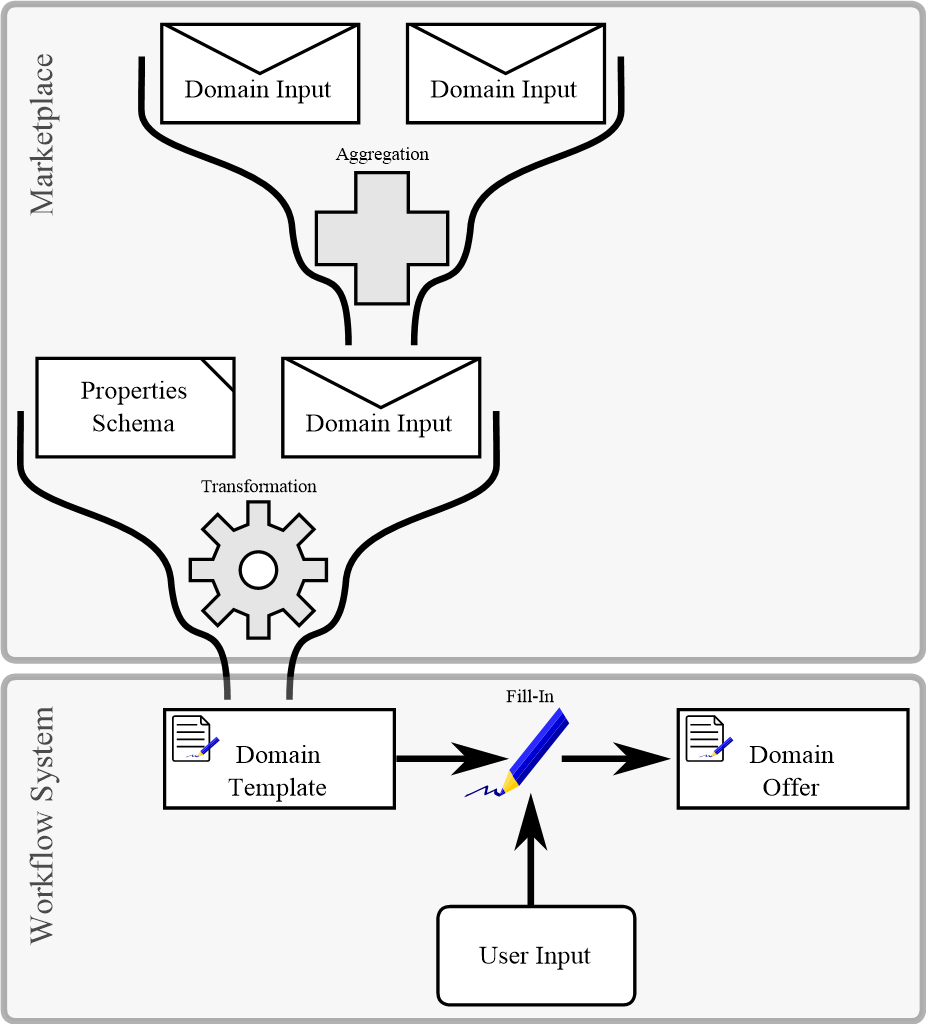}
  \end{center}
  \vspace{-10pt}
  \caption{Class Level Templates: Aggregation and Transformation of Interfaces}
  \label{fig:template_generation}
\end{figure}

As defined in \cite{vigne_structured_2012} the Marketplace allows to define
class level operations (microflows) that in turn invoke instance level
operations (microflows). Invocation does not mean that the microflows are
executed by the Marketplace, but can be fetched from the marketplace in order
to execute them in a Workflow Engine (WFE) at the service consumer's side.

But for class level operations it is also allowed to invoke other class level
operations in order to e.g. realize a \textit{search \& book} operation, which
is the fusion of existing operations.  As the domain designer has full control
over the class level (API), it is not possible that semantically identical
parameters with different names may be used (except as an error by the domain
designer). Thus aggregating WSAG is straight forward, pieces of information can
just be concatenated, duplicate information can be removed.

As properties are valid at class level for all operations, when concatenating
the pieces, identifiers have to be prefixed.

\begin{lstlisting}[caption=Class Level Template,label=lst:template]
<wsag:Template xmlns:xsi="..." xmlns:xs="..."
    xmlns:wsag="..." xsi:schemaLocation="..."
    xmlns:prop="..." xmlns:msg="...">

  <wsag:Name>Cinemas Domain: Search and Book</wsag:Name>
  <wsag:Context/><!-- contains general information -->
  <wsag:CreationConstraints>
    <wsag:Item wsag:name="Zip">
      <wsag:Location>
      //wsag:ServiceDescriptionTerm/prop:address/prop:ZIP
      </wsag:Location>
      <xs:minInclusive xs:value="1000"/>
      <xs:maxInclusive xs:value="9999"/>
    </wsag:Item>
    <wsag:Item wsag:name="DateOfShow">
      <wsag:Location>
      //wsag:ServiceDescriptionTerm/msg:date
      </wsag:Location>
    </wsag:Item>
  </wsag:CreationConstraints>
  <!-- additionl generated information is palced here -->
</wsag:Template>  
\end{lstlisting}

This results in a simple concatenation algorithm depicted in Listing
\ref{lst:template}. As already described in Subsection
\ref{sec:agreements:concepts}, the aggregated WSAG template can then be filled
out by a user and utilized to make offers for a subset of services.

\section{Case Study}
\label{sec:negotiation}

In this Section we justify our approach by analyzing a set of negotiation
protocols. As already mentioned in the introduction, negotiation is the process
of establishing business relations between service providers and consumers. In
the context of this paper this means that a WS-Agreement (WSAG) can be derived,
which can be used in order to call services. Existing class level operations
can be extended to include the functionality to:

\begin{itemize}

  \fitem{(1)} Collect WSAG templates by invoking instance level operations.

  \fitem{(2)} Present aggregated templates to users, by utilizing callbacks
  that can be injected into class level operations (microflows) as presented in
  \cite{vigne_structured_2012}. It is important to note that all templates will
  have the same structure / terms included, but may contain different WSAG
  creation constraints.

  \fitem{(3)} Present offers to services by invoking instance level
  operations with filled out WSAG templates.

  \fitem{(4)} Enact service calls as a result of accepted offers.

\end{itemize}

The instance level operations (microflows) mentioned in (1) and (3) can contain
any kind of calls to actual services. This allows for a high degree of
flexibility: (a) for WSAG unaware services, the logic to create a WSAG could be
either embedded in a microflow, or (b) for WASG aware services results can be
just transformed.

Furthermore, as described in (2) the service consumer (user) can select a
subset of templates that match his preferences. As the service consumer (user)
has also an overview of its past decisions, it can employ strategic or tactic
negotiation as described in the related work. The same goes for service
providers as described in (3).

Because class level operations interact with all participating service
providers, its microflow represents a \textbf{multilateral negotiation
protocol}. A WSAF complaint class level operation has to define a WSAG offer
(\textit{domain offer}) as an input and zero to many WSAG agreements
(\textit{domain agreements}) as output.

On the other hand passing a particular WSAG offer to an instance level
operation for a particular service, can be considered to be an example of a
\textbf{bilateral negotiation protocol}.

As already mentioned in the introduction, this separation between class and
instance level is perfectly suited to map arbitrary per service negotiation
protocols in a transparent fashion. The service consumer can just fetch and
execute microflows, without having to worry about transforming input or
results.

To further exemplify the invoked principles, we evaluated several common
concepts in more detail.

\subsection{Bilateral Negotiation Protocols} 
\label{sec:negotiation:bilateral}

Bilateral negotiation protocols represent how a service consumer is able to
negotiate with a \textbf{specific service provider (one-to-one)}. As for all
negotiations, at the beginning is a \textit{service offer (SO)} and at the end
there is either a \textit{service agreement (SA)} or a cancellation of the
negotiation. What happens in between is covered by the negotiation protocol
(e.g. \textit{send offer}, \textit{confirm agreement} or \textit{quit
negotiation}). In our concept, bilateral negotiation protocols are always
defined using instance level operations.  This is completely in line with their
original intent to provide service specific information to the service consumer
which is also considered as one-to-one information.

In the following, we illustrate our concept with two commonly used bilateral
negotiation protocols.

\subsubsection{Contract Net Interaction Protocol}
\label{sec:negotiation:bilateral:cnip}

\begin{figure}[ht]
  \begin{center}
     \includegraphics[width=0.9\columnwidth]{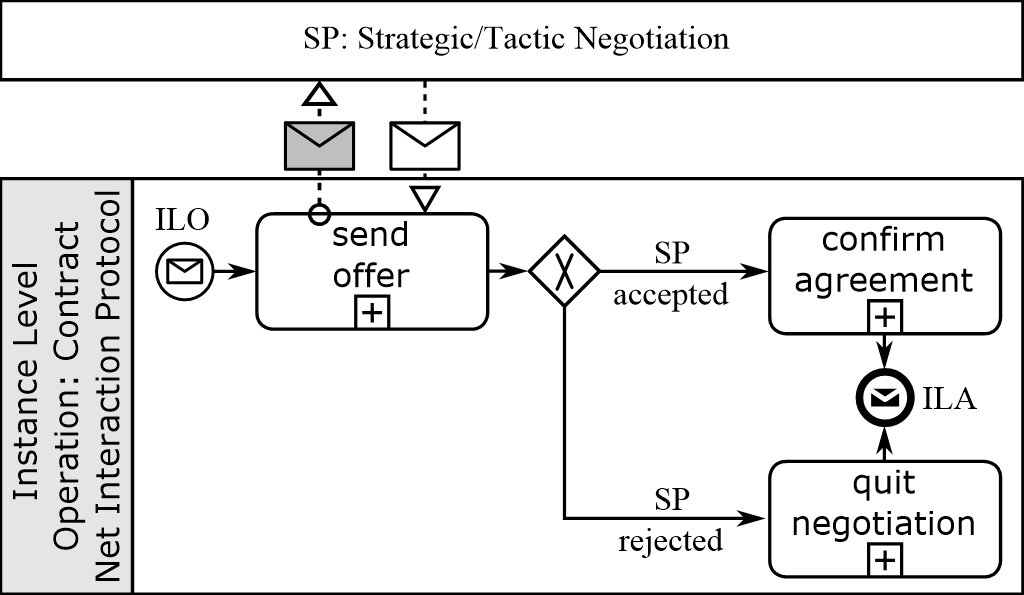}
  \end{center}
  \vspace{-10pt}
  \caption{Bilateral Negotiation: Contract Net Interaction Protocol \cite{smith_contract_1980}}
  \label{fig:negotiation:contract_net}
\end{figure}

The Contract Net Interaction Protocol was introduced by Smith
\cite{smith_contract_1980} in 1980 and represents the protocol which is today
often referred to as ``Supermarket Approach''. This means that the service
consumer offers a contract (\textit{send offer}) for the requested
functionality and the service provider either accepts this offer or rejects it,
without providing any counteroffer (``take it or leave it'').  Because there is
no counteroffer provided for the service consumer, its strategic negotiation
service is not necessary. As an example how a microflow, representing this
protocol in its simplest form (without any parameter transformation), could
look like see Figure  \ref{fig:negotiation:contract_net}.

We decided to include this protocol in our case study as it is  widely used by
today's real world services (e.g.  online-shops, cinema service providers,
\ldots). Furthermore is it referred in a lot of today's research (e.g.
\cite{zulkernine_adaptive_2011, hasselmeyer_implementing_2007,
pichot_dynamic_2008, andrieux_web_2004}) and thus relevant for the Marketplace.

\subsubsection{Alternate Offers Protocol}
\label{sec:negotiation:bilateral:alternate_offer}

\begin{figure}[ht]
  \begin{center}
     \includegraphics[width=0.9\columnwidth]{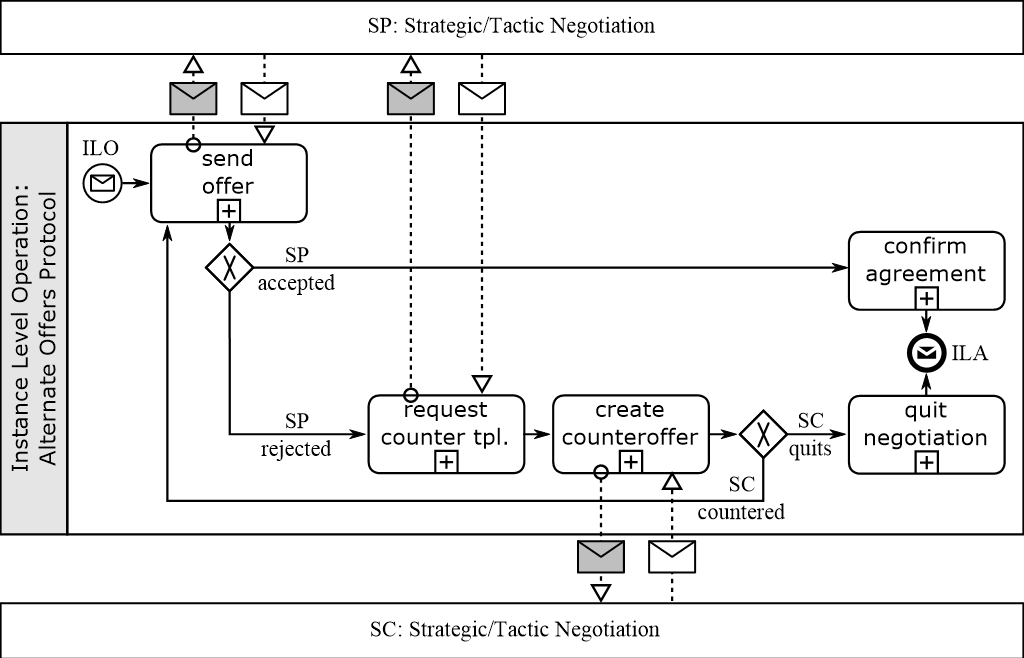}
  \end{center}
  \vspace{-10pt}
  \caption{Bilateral Negotiation: Alternate Offers Protocol \cite{rubinstein_perfect_1982, venugopal_negotiation_2008}}
  \label{fig:negotiation:alternate_offers}
\end{figure}

The ``Alternate Offers Protocol'' was defined by Rubinstein
\cite{rubinstein_perfect_1982} in 1982 and provides the possibility for service
providers (SP) and service consumers (SC) to bargain about the delivered
functionality and usage conditions. As illustrated in Figure
\ref{fig:negotiation:alternate_offers} the protocol allows both parties to
propose counteroffers (\textit{SC: create counteroffer} and \textit{SP: request
counter template}) if one is not willing to accept the current offer.  In WSAG,
each term defines permissible values \cite{andrieux_web_2004} used as
constraints when creating the counteroffer. This cycle goes on until both
parties either accept the current offer or one party quits the negotiation
process.

Because this protocol allows counteroffers, the service consumer must provide a
\textit{strategic negotiation} logic, to decide how the negotiation should
continue and create counteroffers (see Section \ref{sec:negotiation:strategic}
for details). Doing so allows each service consumer to realize his own
bargaining strategy without customizing the instance level operation defined by
the service provider.

Although, as to our best knowledge, not much of today's real world services
support this protocol, it is quite popular among researchers (e.g.
\cite{chhetri_coordinated_2006, comuzzi_architecture_2005,
yan_autonomous_2007}). We agree with Rubinstein et al.
\cite{rubinstein_perfect_1982} that bargaining is essential for the creation of
mature markets.  

\subsection{Multilateral Negotiation Protocols} 
\label{sec:negotiation:multilateral}

So far we focused on bilateral negotiation protocols, defining the interaction
between one specific service provider and a service consumer (one-to-one).  In
this Section we extend our focus to multilateral negotiation protocols, as only
these allow a service consumer to use offers from \textbf{competing service
providers (one-to-many)} for its negotiation strategy. As discussed in e.g.
\cite{rubinstein_perfect_1982,broberg_market-oriented_2008}, if this
information is used to create counteroffers, the overall benefits for the
service consumer will significantly increase, because the service consumer is
completely informed about the market.

Multilateral Negotiation is similar to the ``Alternate Offers Protocol'' (see
Section \ref{sec:negotiation:bilateral:alternate_offer}), only this time the
bargaining is defined at \textit{class level instead of instance level}. The
according operation therefore executes the negotiation protocols, defined by
each service provider, every time a new service agreement should be
negotiated\footnote{It can therefore happen that the ``Alternate Offers
Protocol'' is used inside an iterated multilateral negotiation.}. If
re-negotiation with a particular service providers is not desired, class level
logic can easily exclude them. Again, the service consumer provides a
\textit{strategic negotiation} logic, which is in charge of creating
counteroffers representing the preferences of the service consumer.

The definition of such protocols is built on the interaction of class level and
instance level operations. As this interaction can be utilized in various ways,
a lot of room for domain specific customization is left open, e.g. service
enactment \cite{chhetri_coordinated_2006}, service decomposition
\cite{yan_autonomous_2007}.

We decided to implement the ``Iterated Contract Net Interaction Protocol'',
which was standardized by the FIPA \cite{odell_fipa_2002} and extended with
negotiation coordination capabilities by Yan et al. \cite{yan_autonomous_2007}.

\begin{figure}[ht]
  \begin{center}
     \includegraphics[width=0.9\columnwidth]{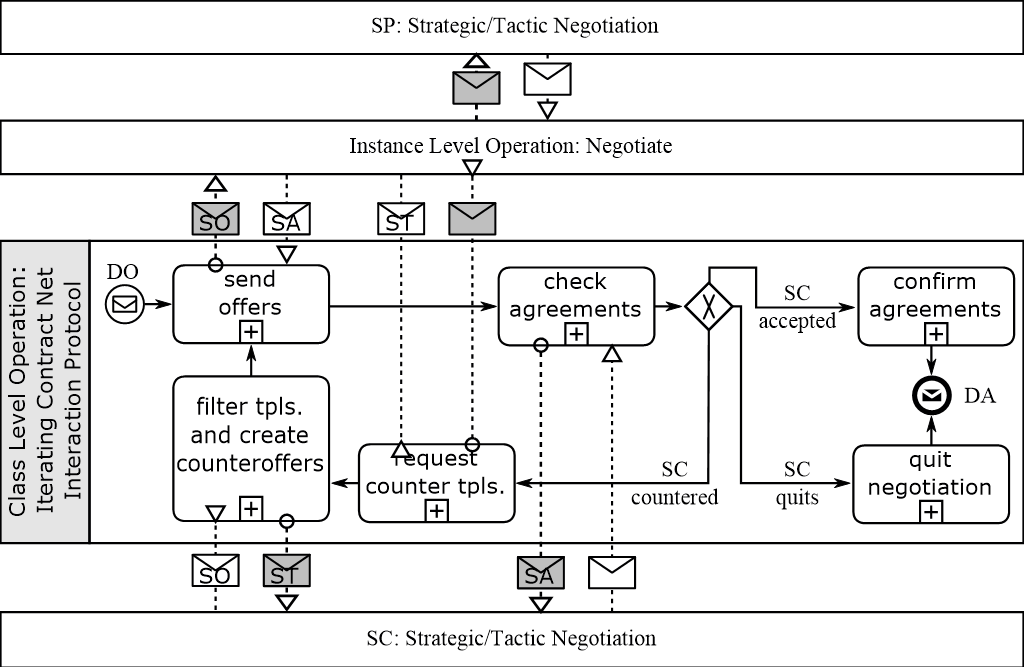}
  \end{center}
  \vspace{-10pt}
  \caption{Multilateral Negotiation: Iterated Contract Net Interaction Protocol \cite{odell_fipa_2002, yan_autonomous_2007}}
  \label{fig:negotiation:competing_resources}
\end{figure}

As illustrated in Figure \ref{fig:negotiation:competing_resources}, during the
negotiation \textit{service offers (SO)} are sent to various service providers
(\textit{send offers}) using the individual bilateral negotiation protocol
(\textit{Instance Level Operation: Negotiate}) of each service provider.
Afterwards the collected \textit{service agreements (SA)} are sent to the
strategic negotiation logic of the service consumer (\textit{check
agreements}). The result of the strategic negotiation is either to accept some
of the agreements (\textit{confirm agreements}), to quit the negotiation
(\textit{quit negotiation}) or to propose counteroffers. If counteroffers are
placed, first the \textit{counter templates (ST)} of the service providers are
collected (\textit{request counter templates}).  Then the \textit{strategic
negotiation} logic of the service consumer filters these templates and creates
counteroffers (\textit{SO}) for all service providers included in the next
iteration (\textit{filter templates and create counteroffers}).

This way of bargaining is also known as ``reverse auction'' where many
different service providers compete. Therefore, in each iteration either the
price decreases or the offered functionality increases.


\subsection{Strategic Negotiation}
\label{sec:negotiation:strategic}

Strategic negotiation is used as a means to enact the service consumers /
providers preferences. Implementing strategic negotiation as part of class /
instance level operations allows to realize individual service preferences
without changing the negotiation protocol provided by the Marketplace and focus
on the outcome of a specific negotiation.

How these preferences are expressed and how competing offers are rated is
strongly affected by micro economics and therefore beyond the focus of this
paper. As pointed out by Faratin et al. \cite{faratin_negotiation_1998}, for
autonomous service negotiation three areas must be considered: (1) Negotiation
Protocols, which are discussed in this case study, (2) Negotiation Issues,
which are covered by  WSAG (see Section \ref{sec:agreements}), and (3)
Negotiation Strategies to define a reasoning model for the service customers
preferences.

They further define that their approach for autonomous negotiation uses a
\textit{value scoring algorithm} as a foundation. In this algorithm, each
single criteria of the negotiation is represented by a \textit{tactic}. Tactics
are functions calculating the value of the assigned criteria. To create an
offer, all tactics are weighted and combined. By varying the weight of each
tactic over time, the \textbf{long-term negotiation strategy} is realized.  As
an example for a negotiation strategy they allege that, in the beginning of
multilateral negotiations, the behavior of competing service providers may
influence the counteroffer more than the value of a single criteria. Therefore
behavior representing tactics are weighted higher than other tactics. However,
at the end of the negotiation phase, single criteria may gain importance. How
tactics and strategy have to be combined to achieve the best negotiation
results, can be educed from various sources e.g. historical data or user's
preferences.

In our approach all this logic is provided on behalf of the service consumer,
thus not directly affects the Marketplace.

\section{Extended Cinemas Example}
\label{sec:example_extended}

\begin{figure}[ht]
  \begin{center}
     \includegraphics[width=1.0\columnwidth]{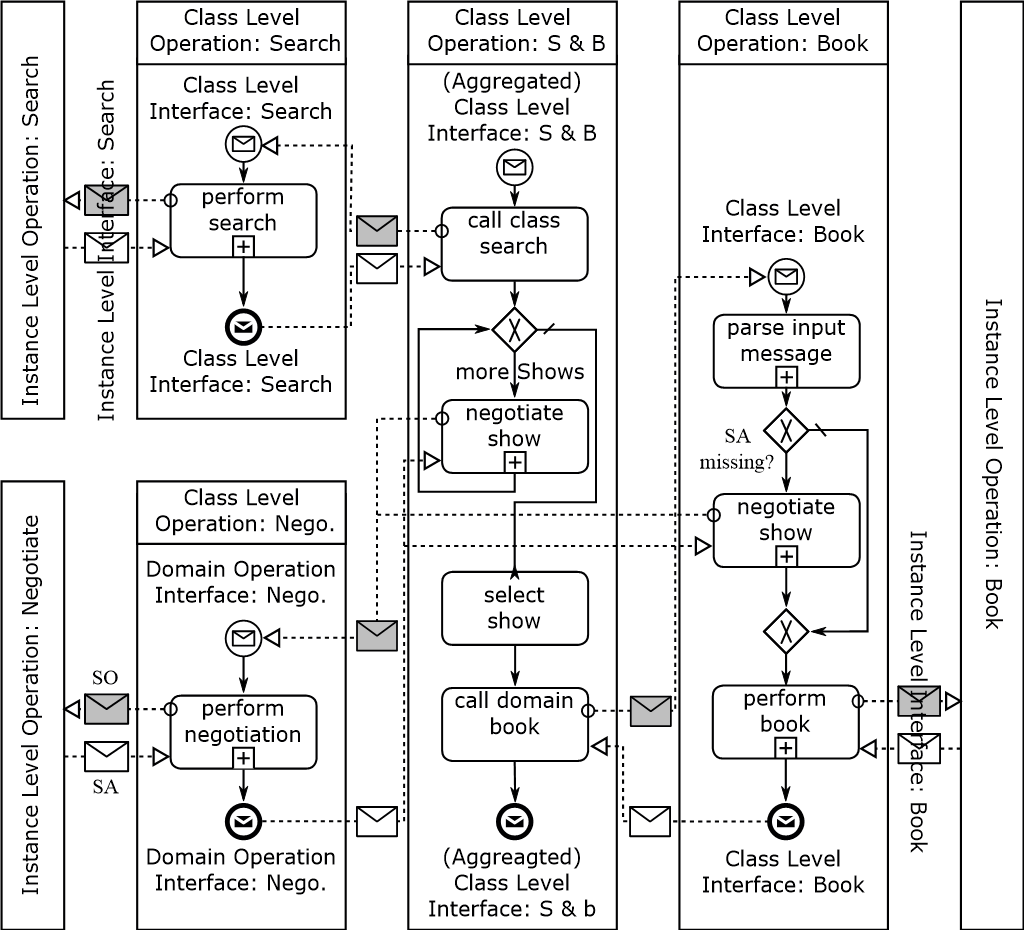}
  \end{center}
  \vspace{-10pt}
  \caption{Multilateral Negotiation: Cinemas Example}
  \label{fig:cinemas_extended}
\end{figure}

So far we introduced our concept for WSAG integration and flexible bilateral
and multilateral negotiation protocols. In this Section we integrate them into
the cinemas example introduced in Section \ref{sec:example} and
\cite{vigne_structured_2012} as illustrated in Figure \ref{fig:cinemas_extended}.

In order to use this operation, the service consumer provides a \textit{domain
offer} for the class level operation \textit{Search \& Book} as only input.

Calling the class level operation \textit{Search} is the same as described in
\cite{vigne_structured_2012}. Only the necessary parameter transformation
before and after the call are different, as the interface of \textit{Search \&
Book} and the internal data representation has changed.

Now that all offered shows are known, negotiations for each show are performed.
To do so the according \textit{class level operation} is called (\textit{negotiate
show}). It is defined inside the \textit{domain interface} of this operation
that a reference to a particular show must be included in the input and
agreements for each show are the output. Using this input message, the
operation is capable to educe the correct service provider and calls its
\textit{instance level operation} for negotiation (\textit{perform negotiation}). Also
parameter transformations to create \textit{service offers (SO)} and to compute
\textit{service agreements (SA)} must be performed.

After all shows are negotiated, the service consumer (acting as the
\textit{strategic negotiation service}) selects the preferred show
(\textit{select show}). Further must all \textit{service agreements}, expect
the selected one, be canceled. This is done as described in the agreement e.g.
sending the agreement to a specific URI. It must be noted that we excluded this
message flow in our illustration to keep it neat.

In the end, the selected \textit{service agreement (SA)} is used as an input
for the class level operation \textit{Book}. In our example this operation defines
in its class level operation interface that either a \textit{service agreement} or a
reference to a particular show is valid input. Depending on this input, the
operation decides (\textit{parse input message}) if it must negotiate the show
(\textit{negotiate show}) before using it or not. In the end the service
operation \textit{Book} of the according service provider can be called.  It
defines in its \textit{instance level operation interface} that a \textit{service
agreement (SA)} must be included to authorize the usage. As output the
\textit{execution results} of the service are define.

This \textit{execution results} and the selected \textit{service agreement}
represent the defined output of \textit{Search \& Book}.

%

\section{Conclusion and Future Work}
\label{sec:conclusion}

In this paper we presented flexible ways to (a) define WSAG contracts for
arbitrary services, and to (b) transparently allow different services to
utilize custom negotiation algorithms and APIs. We showed that both goals could
be achieved by building on an existing Marketplace concept, that promotes
microflows (small pieces of code executable by a Workflow Engine) to achieve a
high level of abstraction for service consumers.

We first described how to extend the concept of unified interfaces to create
unified contracts.  We then introduced how to use the concepts of class
(domain) and instance (service) level to provide flexible negotiation
protocols. Finally we provided a case study to justify our approach.

Within the case study we discussed how (1) bilateral negotiation protocols are
defined using instance level operations, (2) how multilateral negotiation
protocols are defined using class level operations, and (3) how service
consumers can include customized negotiation strategies. We presented how to
implement the ``Contract Net Interaction Protocol'' (because of its prevalence
for real world services) as well as the ``Alternate Offers Protocol'' (as
prerequisite for mature markets). For multilateral negotiation protocols we
presented the ``Iterated Contract Net Interaction Protocol'' in order to
allow various service providers to compete about business. Additionally we
presented one negotiation strategy based on a value scoring algorithm.

The conclusion we can draw from this work is, that by embracing existing
workflow-related concepts, it is possible to build a lightweight solution for
dealing with electronic contracts, in an environment where not all services
support them (or support them in the same way).

Future work will be to exploit the complexity of different types of negotiation
protocols and to implement some custom negotiation strategy services. The
possibilities to monitor complex guarantee terms using class level operations
in the context of service composition is also promising. This can lead to much
improved error handling and service selection, as it may allow for the
prediction of service behavior.





\begin{thebibliography}{10}
\providecommand{\url}[1]{#1}
\csname url@samestyle\endcsname
\providecommand{\newblock}{\relax}
\providecommand{\bibinfo}[2]{#2}
\providecommand{\BIBentrySTDinterwordspacing}{\spaceskip=0pt\relax}
\providecommand{\BIBentryALTinterwordstretchfactor}{4}
\providecommand{\BIBentryALTinterwordspacing}{\spaceskip=\fontdimen2\font plus
\BIBentryALTinterwordstretchfactor\fontdimen3\font minus
  \fontdimen4\font\relax}
\providecommand{\BIBforeignlanguage}[2]{{%
\expandafter\ifx\csname l@#1\endcsname\relax
\typeout{** WARNING: IEEEtran.bst: No hyphenation pattern has been}%
\typeout{** loaded for the language `#1'. Using the pattern for}%
\typeout{** the default language instead.}%
\else
\language=\csname l@#1\endcsname
\fi
#2}}
\providecommand{\BIBdecl}{\relax}
\BIBdecl

\bibitem{ludwig_ws-agreement_2006}
H.~Ludwig, ``{WS-Agreement} concepts and {Use–Agreement-Based}
  {Service-Oriented} architectures,'' \emph{{IBM} Research Division}, 2006.

\bibitem{andrieux_web_2004}
A.~Andrieux, K.~Czajkowski, A.~Dan, K.~Keahey, H.~Ludwig, T.~Nakata, J.~Pruyne,
  J.~Rofrano, S.~Tuecke, and M.~Xu, ``Web services agreement specification
  {(WS-Agreement)},'' in \emph{Global Grid Forum}, 2004.

\bibitem{broberg_market-oriented_2008}
J.~Broberg, S.~Venugopal, and R.~Buyya, ``Market-oriented grids and utility
  computing: The state-of-the-art and future directions,'' \emph{Journal of
  Grid Computing}, vol.~6, no.~3, p. 255–276, 2008.

\bibitem{sakellariou_flexibility_2005}
R.~Sakellariou and V.~Yarmolenko, ``On the flexibility of {WS-agreement} for
  job submission,'' in \emph{Proceedings of the 3rd international workshop on
  Middleware for grid computing}, 2005, p. 1–6.

\bibitem{haq_framework_2009}
I.~Ul~Haq, A.~Huqqani, and E.~Schikuta, ``Aggregating hierarchical service
  level agreements in business value networks,'' in \emph{Business Process
  Management: 7th International Conference, BPM 2009, Ulm, Germany, September
  8-10, 2009. Proceedings 7}.\hskip 1em plus 0.5em minus 0.4em\relax Springer,
  2009, pp. 176--192.

\bibitem{zulkernine_adaptive_2011}
F.~H. Zulkernine and P.~Martin, ``An adaptive and intelligent {SLA} negotiation
  system for web services,'' \emph{{IEEE} Transactions on Services Computing},
  vol.~4, no.~1, pp. 31--43, Jan. 2011.

\bibitem{yan_autonomous_2007}
J.~Yan, R.~Kowalczyk, J.~Lin, M.~B. Chhetri, S.~K. Goh, and J.~Zhang,
  ``Autonomous service level agreement negotiation for service composition
  provision,'' \emph{Future Generation Computer Systems}, vol.~23, no.~6, pp.
  748--759, Jul. 2007.

\bibitem{ludwig_cremona:_2004}
H.~Ludwig, A.~Dan, and R.~Kearney, ``Cremona: an architecture and library for
  creation and monitoring of {WS-agreents},'' in \emph{Proceedings of the 2nd
  International Conference on Service oriented computing}, 2004, p. 65–74.

\bibitem{venugopal_negotiation_2008}
S.~Venugopal, X.~Chu, and R.~Buyya, ``A negotiation mechanism for advance
  resource reservations using the alternate offers protocol,'' in \emph{Quality
  of Service, 2008. {IWQoS} 2008. 16th International Workshop on}, 2008, p.
  40–49.

\bibitem{comuzzi_architecture_2005}
M.~Comuzzi and B.~Pernici, ``An architecture for flexible web service qos
  negotiation,'' in \emph{Ninth IEEE International EDOC Enterprise Computing
  Conference (EDOC'05)}.\hskip 1em plus 0.5em minus 0.4em\relax IEEE, 2005, pp.
  70--79.

\bibitem{parkin_framework_2006}
M.~Parkin, D.~Kuo, and J.~Brooke, ``A framework and negotiation protocol for
  service contracts,'' in \emph{2006 {IEEE} International Conference on
  Services Computing {(SCC'06)}}, Chicago, {IL}, {USA}, Sep. 2006, pp.
  253--256.

\bibitem{hasselmeyer_implementing_2007}
P.~Hasselmeyer, H.~Mersch, B.~Koller, H.~N. Quyen, L.~Schubert, and P.~Wieder,
  ``Implementing an {SLA} negotiation framework,'' in \emph{Proceedings of the
  {eChallenges} e-2007 Conference, Hague, Netherlands}, 2007.

\bibitem{pichot_dynamic_2008}
A.~Pichot, P.~Wieder, O.~Wäldrich, and W.~Ziegler, ``Dynamic {SLA-negotiation}
  based on {WS-Agreement},'' in \emph{Proceedings of the 4th International
  conference on Web Information Systems and Technologies {(WEBIST} 2008)},
  2008, p. 38–45.

\bibitem{faratin_negotiation_1998}
P.~Faratin, C.~Sierra, and N.~R. Jennings, ``Negotiation decision functions for
  autonomous agents,'' \emph{Robotics and Autonomous Systems}, vol.~24, no.
  3-4, p. 159–182, 1998.

  \newpage
  
\bibitem{odell_fipa_2002}
\BIBentryALTinterwordspacing
J.~Odell, S.~Poslad, and R.~Levy, ``{FIPA} iterated contract net interaction
  protocol specification,'' Dec. 2002. [Online]. Available:
  \url{http://www.fipa.org/http://fipa.org/specs/fipa00030/SC00030H.pdf}
\BIBentrySTDinterwordspacing

\bibitem{vigne_structured_2012}
R.~Vigne, J.~Mangler, E.~Schikuta, and S.~Rinderle-Ma, ``A structured
  marketplace for arbitrary services,'' \emph{Future Generation Computer
  Systems}, vol.~28, no.~1, pp. 48--57, 2012.


\bibitem{schikuta1998vipios}
E.~Schikuta, T.~Fuerle, and H.~Wanek, ``Vipios: The vienna parallel
  input/output system,'' in \emph{Euro-Par’98 Parallel Processing: 4th
  International Euro-Par Conference}.\hskip 1em plus 0.5em minus 0.4em\relax
  Springer, 1998, pp. 953--958.

\bibitem{mach2012generic}
W.~Mach and E.~Schikuta, ``A generic negotiation and re-negotiation framework
  for consumer-provider contracting of web services,'' in \emph{Proceedings of
  the 14th International Conference on Information Integration and Web-based
  Applications \& Services}, 2012, pp. 348--351.

\bibitem{schikuta2004n2grid}
E.~Schikuta and T.~Weishaupl, ``N2grid: neural networks in the grid,'' in
  \emph{2004 IEEE International Joint Conference on Neural Networks (IEEE Cat.
  No. 04CH37541)}, vol.~2.\hskip 1em plus 0.5em minus 0.4em\relax IEEE, 2004,
  pp. 1409--1414.

\bibitem{cs745}
\BIBentryALTinterwordspacing
E.~Schikuta, F.~Donno, H.~Stockinger, H.~Wanek, T.~Weish{\"a}upl, E.~Vinek, and
  C.~Witzany, ``Business in the grid: Project results,'' in \emph{1st Austrian
  Grid Symposium}.\hskip 1em plus 0.5em minus 0.4em\relax OCG, December 2005.
  [Online]. Available: \url{http://eprints.cs.univie.ac.at/745/}
\BIBentrySTDinterwordspacing

\bibitem{schikuta2008grid}
E.~Schikuta, H.~Wanek, and I.~Ul~Haq, ``Grid workflow optimization regarding
  dynamically changing resources and conditions,'' \emph{Concurrency and
  Computation: Practice and Experience}, vol.~20, no.~15, pp. 1837--1849, 2008.

\bibitem{kofler2009parallel}
K.~Kofler, I.~ul~Haq, and E.~Schikuta, ``A parallel branch and bound algorithm
  for workflow qos optimization,'' in \emph{2009 International Conference on
  Parallel Processing}.\hskip 1em plus 0.5em minus 0.4em\relax IEEE, 2009, pp.
  478--485.

\bibitem{stuermer2009building}
G.~Stuermer, J.~Mangler, and E.~Schikuta, ``Building a modular service oriented
  workflow engine,'' in \emph{2009 IEEE international conference on
  service-oriented computing and applications (SOCA)}.\hskip 1em plus 0.5em
  minus 0.4em\relax IEEE, 2009, pp. 1--4.

\bibitem{oldham_semantic_2006}
N.~Oldham, K.~Verma, A.~Sheth, and F.~Hakimpour, ``Semantic {WS-agreement}
  partner selection,'' in \emph{Proceedings of the 15th international
  conference on World Wide Web}, 2006, p. 697–706.

\bibitem{kuebert_restful_2011}
R.~Kübert, G.~Katsaros, and T.~Wang, ``A {RESTful} implementation of the
  {WS-Agreement} specification,'' in \emph{Proceedings of the Second
  International Workshop on {RESTful} Design}, 2011, p. 67–72.

\bibitem{zulkernine_policy-based_2009}
F.~Zulkernine, P.~Martin, C.~Craddock, and K.~Wilson, ``A {Policy-Based}
  middleware for web services {SLA} negotiation,'' in \emph{{IEEE}
  International Conference on Web Services, 2009. {ICWS} 2009}.\hskip 1em plus
  0.5em minus 0.4em\relax {IEEE}, Jul. 2009, pp. 1043--1050.

\bibitem{smith_contract_1980}
R.~G. Smith, ``The contract net protocol: High-level communication and control
  in a distributed problem solver,'' \emph{Computers, {IEEE} Transactions on},
  vol. 100, no.~12, p. 1104–1113, 1980.

\bibitem{hung2004ws}
P.~C. Hung, H.~Li, and J.-J. Jeng, ``Ws-negotiation: an overview of research
  issues,'' in \emph{37th Annual Hawaii International Conference on System
  Sciences, 2004. Proceedings of the}.\hskip 1em plus 0.5em minus 0.4em\relax
  IEEE, 2004, pp. 10--pp.

\bibitem{rubinstein_perfect_1982}
A.~Rubinstein, ``Perfect equilibrium in a bargaining model,''
  \emph{Econometrica: Journal of the Econometric Society}, p. 97–109, 1982.

\bibitem{altmann_vieslaf_2009}
\BIBentryALTinterwordspacing
I.~Brandic, D.~Music, P.~Leitner, and S.~Dustdar, ``{VieSLAF} framework:
  Enabling adaptive and versatile {SLA-Management},'' in \emph{Grid Economics
  and Business Models}, J.~Altmann, R.~Buyya, and O.~F. Rana, Eds.\hskip 1em
  plus 0.5em minus 0.4em\relax Berlin, Heidelberg: Springer Berlin Heidelberg,
  2009, vol. 5745, pp. 60--73. [Online]. Available:
  \url{http://www.springerlink.com/content/l380u04815284287/}
\BIBentrySTDinterwordspacing

\bibitem{chhetri_coordinated_2006}
M.~B. Chhetri, J.~Lin, S.~K. Goh, J.~Yan, J.~Y. Zhang, and R.~Kowalczyk, ``A
  coordinated architecture for the agent-based service level agreement
  negotiation of web service composition,'' in \emph{Software Engineering
  Conference, 2006. Australian}, 2006, p. 10–pp.

\end{thebibliography}
%
\end{document}